# Observation of fractality-induced topology in photonic crystals

Bei Yan[1,2,#], Yingfeng Qi[2,#], Xiang Xi[3], Linyun Yang[4], Yan Meng[3], Zhen-Xiao Zhu[5], Jing-Ming Chen[2], Ziyao Wang[2], Zhen Gao[2,*]

*[1] Hubei Province Key Laboratory of Systems Science in Metallurgical Process, and School of Physics and Mechanics, Wuhan University of Science and Technology, Wuhan 430081, China*
*[2] State Key Laboratory of Optical Fiber and Cable Manufacturing Technology, Department of Electronic and Electrical Engineering, Guangdong Key Laboratory of Integrated Optoelectronics Intellisense, Southern University of Science and Technology, Shenzhen 518055, China*
*[3]School of Electrical Engineering and Intelligentization, Dongguan University of Technology, Dongguan, 523808, China*
*[4]College of Aerospace Engineering, Chongqing University, Chongqing, 400030, China*
*[5]Frontier Interdisciplinary Domain, Beijing Institute of Technology, Zhuhai 519088, China.*
*[#]These authors contributed equally to this work.*
*[*]Corresponding author. Email: gaoz@sustech.edu.cn (Z.G.)*

**Fractal topology—achieved by integrating nontrivial topology into fractal geometries with self-similarity and non-integer dimensions—has opened new avenues for exploring topological phases of matter. Recent theoretical advances revealed a counterintuitive fractal topology: fractality itself can induce nontrivial topology in an otherwise trivial system. Here, we report the first experimental observation of fractality-induced topology in a tight-binding-like photonic crystal, without relying on traditional driving mechanisms such as magnetic fields, staggered hopping, or spin-orbit coupling. We demonstrate that fractality alone is sufficient to lift the degeneracy of Kagome lattice band structure and induce topological corner states within the bandgap of the resulting fractal Kagome photonic crystal, which is a photonic higher-order topological insulator. This work experimentally reveals a novel mechanism for realizing nontrivial topological states, expanding both the fundamental frontier and potential application of topological physics.**

## Introduction

The discovery of symmetry-protected topological phases of matter [1,2] has fundamentally reshaped our understanding of quantum matter in condensed-matter physics and inspired the emergence of topological states in classical-wave systems, such as photonics [3-6], acoustics [7,8], mechanics [9,10], electric circuits [11-13], and cold atoms [14,15]. A hallmark feature of topological phases is the robust topological states that arise from the intricate interplay between a system's nontrivial bulk topology and its underlying symmetries, as described by the celebrated bulk-boundary correspondence. A quintessential example is the quantum Hall effect [16-18], whose nontrivial band topology is driven by magnetic fields that break time-reversal symmetry (TRS). This was followed by the quantum spin Hall effect [19-23], a topological phase achieved by strong spin-orbit coupling without breaking TRS. Subsequently, the concept of symmetry protection was extended to crystalline symmetries (e.g., reflection, inversion, rotation), yielding topological crystalline insulators [24-31] and higher-order topological insulators (HOTIs) [32-43].

Recently, the exploration of topological phases has been extended to fractal geometries—structures defined by self-similarity and non-integer dimensions [44-46]. This moves beyond the canonical bulk-boundary correspondence because fractal geometries, such as the Sierpiński gasket, lack a conventional bulk. Topological states, including higher-order topological phases, have been successfully implemented on such fractal lattices across various systems [47-59]. However, these demonstrations have predominantly involved projecting a pre-existing, topologically non-trivial model onto a fractal framework, rather than generating topology from the fractal geometry itself. Strikingly, L. Eek et al. [60] theoretically predicted that nontrivial topology can indeed be induced by the fractal structure itself, even without traditional driving mechanisms such as magnetic fields, spin-orbit coupling, or staggered hopping. This finding goes beyond traditional paradigms of topological origin and opens new avenues for topological material design by leveraging intrinsic self-similarity. Yet, to date, experimental observation of this fractality-induced topology remains elusive.

In this work, we report the first experimental observation of a fractality-induced photonic HOTI in a tight-binding-like Kagome photonic crystal with Sierpiński gasket fractal geometry. Using microwave near-field measurements and Fourier transform analysis, we demonstrate experimentally that the intrinsic K-point degeneracy of the Kagome lattice—with equal inter- and intra-cell couplings—can be lifted, and a complete photonic bandgap can be opened, solely by the fractal geometry. Furthermore, we directly observe topological corner states within the fractality-induced photonic bandgap, in excellent agreement with theoretical predictions and numerical simulations. Our results establish fractality itself as a novel mechanism for generating topologically protected photonic states, paving the way for the design of advanced topological devices with non-integer dimensions.

**Results**

We begin with a Kagome lattice with equal inter- and intra-cell couplings, $w = v = 1$, whose unit cell and gapless bulk band structure, featuring a degenerate Dirac cone at the K point, are shown in Figs. 1a and 1b, respectively (see the Hamiltonian and band structures in Supplementary Material Note 1). By introducing fractality (Sierpiński gasket fractal geometries) into the Kagome lattice, we obtain a first-generation Sierpiński fractal Kagome lattice, as shown in Fig. 1c. The fractal lifts the degenerate Dirac cone at the K point and generates a bandgap within which the topological corner states emerge, as shown in Figs. 1d and 1e (see more details in Supplementary Material Note 2). For the Kagome lattice, it is well known that realizing topological states conventionally requires the introduction of breathing coupling, in which intra- and inter-cell couplings satisfy $v < w$. In contrast, the self-similar nature of the Sierpiński fractal Kagome lattice enables the introduction of isospectral reduction (ISR). This is achieved by partitioning the nine lattice sites of the Sierpiński fractal Kagome lattice (described by Hamiltonian $\mathcal{H}_1$) into two subsets: subset S (yellow sites 1–3) and its complement $\overline{\mathrm{S}}$ (blue sites 4–9), as shown in Fig. 1e. The ISR of $\mathcal{H}_1$ is then defined as: $\mathcal{R}_S(\mathcal{H}_1, E) \equiv (H_{SS} + H_{S\bar{S}}[H_{S\bar{S}} - EI]^{-1}H_{\bar{S}S})$, following the methodology in Ref. [61–66]. This reduction transforms the original Hamiltonian $\mathcal{H}_1$ of the Sierpiński fractal

Kagome lattice into an effective breathing Kagome lattice Hamiltonian $\mathcal{R}_S(\mathcal{H}_1, E)$. In this reduced lattice, the inter-cell coupling is fixed at $w = 1$, while the intra-cell coupling $v(E)$ and the onsite potential $a(E)$ become energy-dependent functions given by $v(E) = E/[(E^2-1)(E-2)]$ and $a(E) = 2(E^2-E-1)/[(E^2-1)(E-2)]$, respectively, which vary with the eigenenergy $E$ of the original Sierpiński fractal Kagome lattice Hamiltonian (see more details in Supplemental Material Note 3). As shown in Fig. 1f, the equivalent breathing Kagome lattice, whose intra-cell coupling is smaller than the inter-cell coupling ($v(E) < w$), supports topological corner states at eigenenergy $E = -0.414$, which is exactly the eigenenergy of the topological corner states of the Sierpiński fractal Kagome lattice [Fig. 1e], indicating that fractality itself can induce higher-order band topology.

To implement the fractality-induced HOTI in a photonic system, we adopt the tight-binding-like Kagome photonic crystal exhibiting scalar-wave-like band structures [67-70], as shown in the left panel of Fig. 2a, in which the dielectric rods (grey dots) serve as sites, and the metallic rods (brown dots) are introduced to confine the Mie resonance to satisfy the tight-binding approximation in photonic systems. By introducing fractality into the Kagome photonic crystal, we can achieve the first- and second-generation of Sierpiński fractal Kagome photonic crystals, as shown in the upper-right and lower-right panels of Fig. 2a. Figure 2b shows the simulated bulk band structure of the Kagome photonic crystal with equal inter-cell and intra-cell couplings, which exhibits a Dirac point at the K point. As the first- and second-generation fractality is introduced, the degeneracy of the Dirac point is lifted and opens three (seven) topological (yellow region) and one (four) trivial (grey region) photonic band gaps for the first-generation (second-generation) Sierpiński fractal Kagome photonic crystal, as shown in Figs. 2c and 2d, respectively. The topological or trivial band gaps are determined by the symmetry eigenvalues of the photonic bands at the high-symmetry points (HSPs). Given that the Sierpiński fractal Kagome photonic crystals exhibit $C_3$ rotational symmetry, topological (trivial) band gaps will (will not) exhibit a discrepancy in the number of bands with the rotational eigenvalue at different HSPs, which is denoted by rotational invariant $\chi^{(3)}$ = ($[K_1^{(3)}]$, $[K_2^{(3)}]$) and $\left[\mathrm{K}_p^{(3)}\right] \equiv \#\mathrm{K}_p^{(3)} - \#\Gamma_p^{(3)}$, where $\#\mathrm{K}_p^{(3)}$ and $\#\Gamma_p^{(3)}$ is the number of bands below the band gaps with rotational eigenvalue at HSPs K and Γ, $p$ = 1, 2 [38] (see more details in Supplemental Material Note 4). Topological corner states emerge within the topological bandgaps with nontrivial $\chi^{(3)} \neq 0$ and nontrivial nominal electronic corner charge $Q_{corner}^{(3)} = \left[K_2^{(3)}\right] \neq 0$. We then design two finite first- and second-generation Sierpiński fractal Kagome photonic crystals and calculate their eigenstate spectra, as shown in Figs. 2e and 2g, respectively. We use colored dots to distinguish topological corner states in different topological band gaps, and their eigenmode distributions are shown in Figs. 2f and 2h. There are three degenerate topological corner states (blue, red, and green dots) in each topological bandgap, whose field distributions are localized at three different corners of the Sierpiński fractal Kagome photonic crystal. Note that based on the topological band gaps (indicated by the yellow regions) shown in Figs. 2c and 2d, a greater number of topological corner states should be observable. However, due to the extremely narrow band gap, the other topological corner states overlap with bulk states, making them difficult to distinguish.

To experimentally demonstrate the fractality-induced photonic HOTI, we fabricate three tight-binding-like Kagome photonic crystals comprising triangular air foam embedded with dielectric (black circles) and metallic (copper color) rods, as shown in Figs. 3a-3c, the upper metallic plate is removed to see the inner structures. The experimental setup consists of a vector network analyzer (Keysight E5080) and two monopole antennas: one as a point source to excite the composite structure, and the other as a probe inserted into the sample to measure the transmission spectra and $E_z$ field distributions. We then perform a Fast Fourier transform on the measured $E_z$ field distributions to extract the experimental photonic band structures (color maps), which agree well with the simulated results (white solid lines), as shown in Figs. 3d-3f. For the Kagome photonic crystal shown in Fig. 3a, the distances between the intra-cell and inter-cell dielectric rods are the same; thus, its photonic band structure exhibits a symmetry-protected Dirac point at the K point, as shown in Fig. 3d. After introducing fractals to the Kagome photonic crystal, the Dirac point is lifted, and photonic band gaps appear for the first- and second-generation Sierpiński fractal Kagome photonic crystals, as shown in Figs. 3e and 3f. The experimental results agree well with the simulation results and unambiguously verify that fractality itself can induce topological photonic band gaps.

Next, we experimentally demonstrate the fractality-induced topological corner states in the photonic band gaps. We first place a point source (green star) near the left corner and insert a probe antenna into the center (grey star) of the first-generation Sierpiński fractal Kagome photonic crystal, as illustrated in Fig. 4b. By measuring its bulk transmission spectrum (grey line) shown in Fig. 4a, we observe three transmission dips that correspond to the simulated photonic band gaps shown in Fig. 2e. We then insert a probe antenna at the sample's corner (red star) to measure the transmission spectrum of the topological corner states (red line). We observe that two transmission peaks emerge in the two topological band gaps (yellow region), indicating the existence of topological corner states. While in the trivial bandgap (grey region), there exist no transmission peaks. To directly observe the topological corner states, we insert a probe antenna into the small air holes one by one to map the $E_z$ field distribution of the topological corner states, as shown in Figs. 4c and 4d, respectively. The measured $E_z$ distributions are mainly concentrated at the corner, revealing the tightly localized nature of the topological corner states. The experimental results agree well with the simulation results shown in Fig. 2f, verifying the fractality-induced photonic topological corner states. We also measure the transmission spectra of the topological corner (blue line) and bulk (grey line) states of the second-generation Sierpiński fractal Kagome photonic crystal, as shown in Fig. 4e. The second-generation Sierpiński fractal photonic crystal supports more topological corner states in three topological band gaps (yellow region), and the measured $E_z$ field distributions of the three topological corner states are shown in Figs. 4f-4h, agreeing well with the simulation results shown in Fig. 2h.

## Discussion

In conclusion, we have experimentally realized the first fractality-induced photonic HOTIs in tight-binding-like fractal Kagome photonic crystals. We demonstrate that fractals themselves can give rise to topological photonic band gaps in a trivial system and, within these gaps, induce localized topological corner states. Moreover, we show that a higher-generation fractal can generate more topological band gaps and corner states. These results experimentally establish a new physical mechanism for realizing nontrivial topological states and open a new avenue for the design of robust fractal photonic devices. Looking ahead, the exploration of fractality-induced topological surface and hinge states in three-dimensional systems also holds significant promise. Furthermore, the underlying mechanism is general and can be extended to a variety of other physical platforms, including condensed-matter systems, acoustic and mechanical structures, electric circuits, and cold-atom setups.

## Methods

**Numerical simulations.** All numerical results presented in this work are simulated using the RF module of COMSOL Multiphysics. The bulk band structures are calculated using a unit cell with periodic boundary conditions in all directions. The metallic rods are modeled as perfect electric conductors (PEC) in the simulation. In the full-wave simulations of a finite tight-binding-like Kagome photonic crystal, all boundaries are set as open boundary conditions.

**Materials and experimental setups.** The copper plates are fabricated by depositing a 0.035 mm-thick copper layer onto a Teflon-woven-glass fabric laminate. We use perforated air foam (ROHACELL 31 HF with a relative permittivity of 1.04 and a loss tangent of 0.0025) to fix the metallic and dielectric rods. In the experimental measurements, the amplitude and phase of the electric fields are measured using a vector network analyzer (Keysight E5080) connected by two electric dipole antennas serving as the source and probe, respectively. To excite the fractality-induced topological corner states, a point source antenna is placed at the corner of the sample, and a probe antenna is inserted into the air holes one by one to scan the electric fields.

**Data availability**

The data that support the findings of this study are available from the corresponding authors upon reasonable request.

**Code availability**

We use commercial software COMSOL Multiphysics to perform electromagnetic numerical simulations. Requests for computation details can be addressed to the corresponding authors.

**Acknowledgments**

Z.G. acknowledges funding from the National Key R&D Program of China (grant no. 2025YFA1412300), National Natural Science Foundation of China (grant no. 62361166627 and 62375118), Guangdong Basic and Applied Basic Research Foundation (grant no. 2024A1515012770), Shenzhen Science and Technology Innovation Commission (grants no. 20230802205352003), and High-level Special Funds (grant no. G03034K004). B.Y. acknowledges the support from the Hubei Provincial Natural Science Foundation of China under grant No. 2025AFB011. X.X. acknowledges the support from the National Natural Science Foundation of China under grant no. 62405053, and the Basic and Applied Basic Research Foundation of Guangdong Province under grant no. 2025A1515012229. Y.M. acknowledges the support from the National Natural Science Foundation of China under Grant no. 12304484, Guangdong Basic and Applied Basic Research Foundation under grant no. 2024A1515011371. Z.Z. acknowledges the support from the National Natural Science Foundation of China under grant no. 12404053, Guangdong Basic and Applied Basic Research Foundation under grant no. 2025A1515012284. L.Y. acknowledges the support from the National Natural Science Foundation of China under grant no. 12502106, and the Natural Science Foundation of Chongqing under grant no. CSTB2025NSCQ-GPX0776.

**Authors Contributions**

Z.G. initiated and supervised the project. B.Y. performed the simulations. B.Y., Y.F.Q, and Z.G. designed the experiments. B.Y., Z.G., Z.Y.W., Y.F.Q., X.X., L.Y., Y.M., Z.X.Z., and J.M.C. fabricated samples. B.Y., Y.F.Q., and Z.Y.W. carried out the measurements. B.Y., Y.F.Q., Z.Y.W., X.X., and Z.G. analyzed the data. B.Y. and Z.G. wrote the manuscript. Z.G. revised the manuscript.

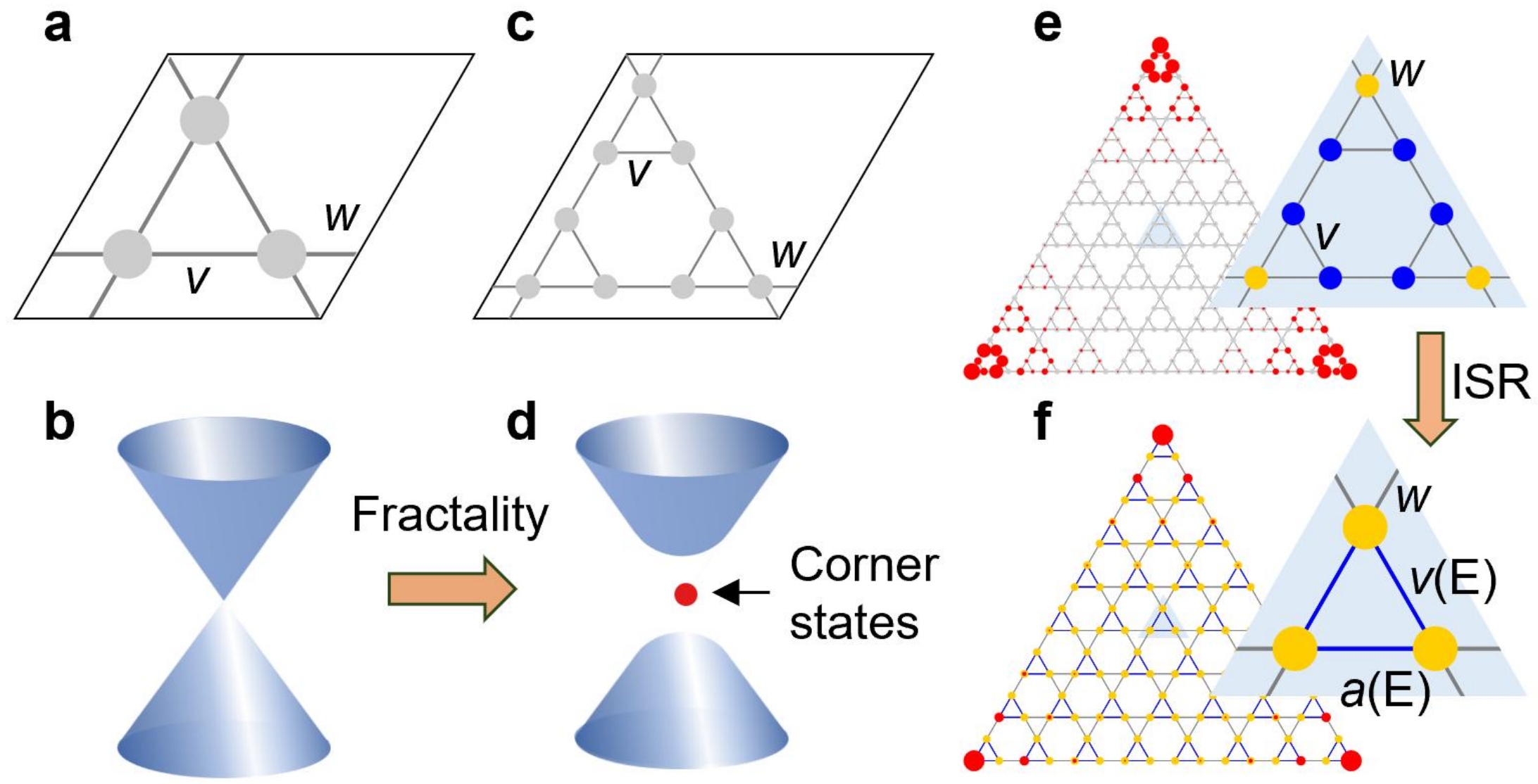

**Fig. 1 | Schematic diagram of the fractality-induced HOTIs**. **a** The unit cell of a Kagome lattice with equal inter- and intra-cell couplings $w = v = 1$. **b** The gapless bulk band structure of the Kagome lattice featuring a degenerate Dirac cone. **c** The unit cell of a Sierpiński fractal Kagome lattice. **d** The gapped bulk band structure of the Sierpiński fractal Kagome lattice. Topological corner states (red dot) emerge in the fractality-induced topological band gap. **e** The energy distribution of the topological corner states of the Sierpiński fractal Kagome lattice at $E = -0.414$. The yellow (blue) dots represent subset S ($\overline{\mathrm{S}}$) **f** The energy distribution of the topological corner state of the equivalent breathing Kagome lattice with intra-cell hopping ($v(E) = 0.414$) smaller than the inter-cell hopping ($w = 1$) and onsite potential $a(E) = -1.414$ at $E = -0.414$.

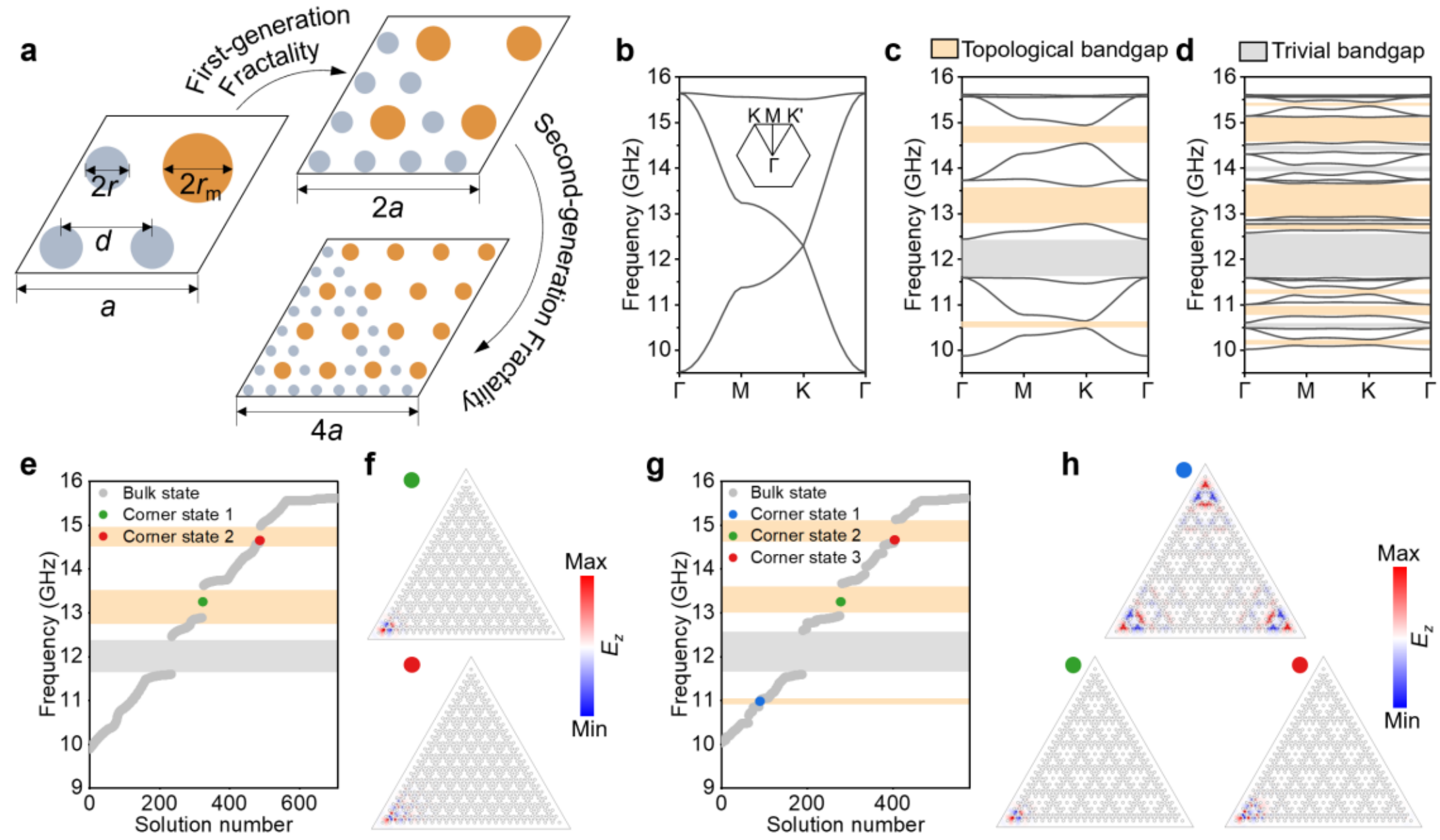

**Fig. 2 | Fractality-induced photonic HOTIs in tight-binding-like fractal Kagome photonic crystals**. **a** The unit cell of the Kagome photonic crystal (left panel) and first- (right-upper panel) and second-generation (right-lower panel) Sierpiński fractal Kagome photonic crystals. The grey (brown) dots represent dielectric (metallic) rods. **b** The simulated bulk band structure of the Kagome photonic crystal. **c**-**d** The simulated bulk band structures of the first- (**c**) and second-generation (**d**) Sierpiński fractal Kagome photonic crystals. The yellow (grey) regions represent topological (trivial) band gaps. **e** Simulated eigenstate spectrum of a finite first-generation Sierpiński fractal Kagome photonic crystal. The bulk (corner) states are represented by grey (green and red) dots, respectively. **f** Simulated electric field distributions of the topological corner states marked in (**e**). **g** Simulated eigenstate spectrum of a finite second-generation Sierpiński fractal Kagome photonic crystal. The bulk (corner) states are represented by grey (blue, green, and red) dots, respectively. **h** Simulated electric field distributions of the topological corner states marked in (**g**).

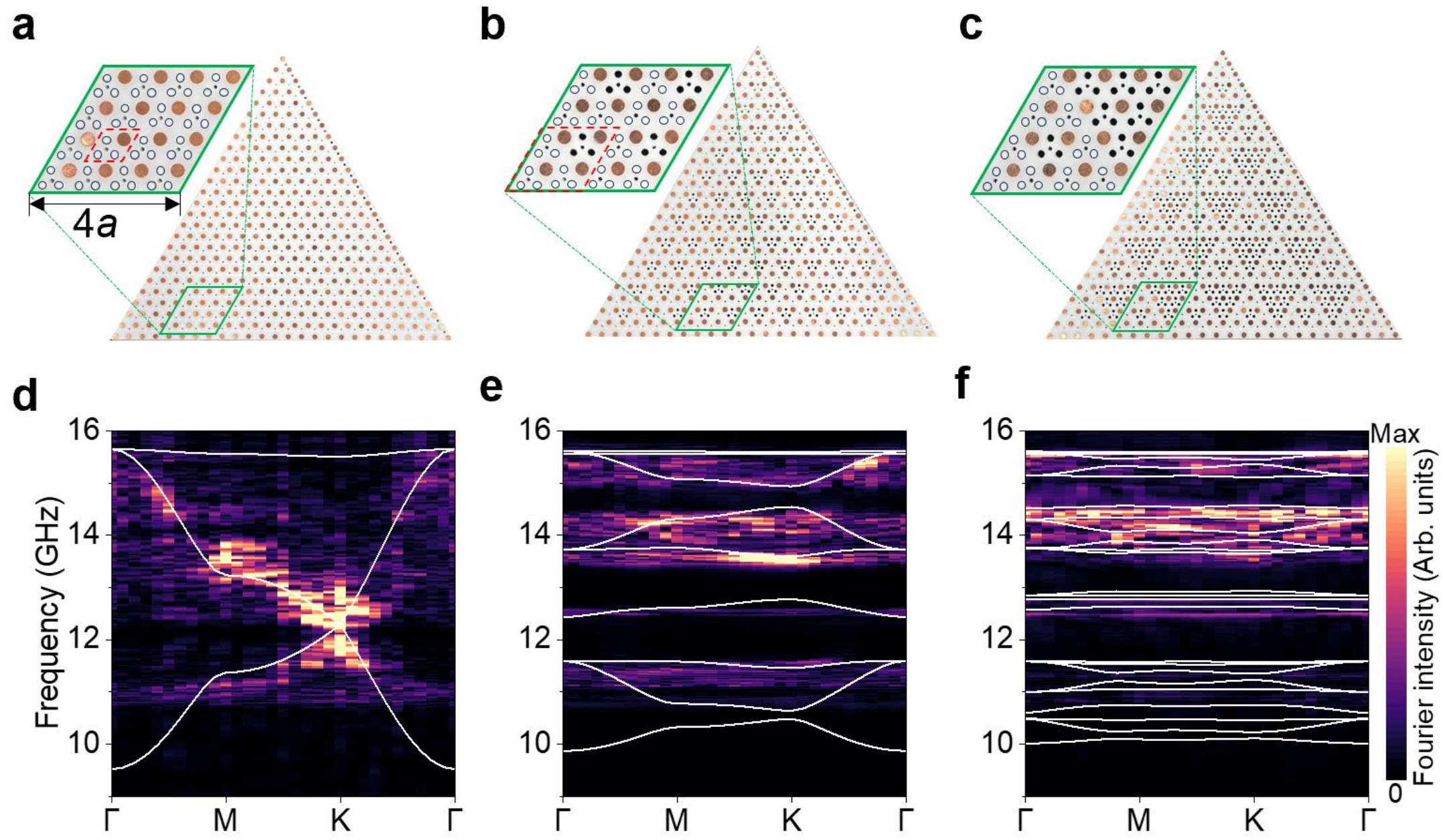

**Fig. 3 | Experimental observation of fractality-induced topological photonic band gaps**. **a** Top view of the fabricated Kagome photonic crystal. The dielectric (black circles) and metallic (copper color) rods are inserted in a triangular air foam. The upper-left inset shows an enlarged view of 4×4 unit cells (the red dashed rhomboid represents a unit cell). The top metallic plate is removed to see the inner structure. **b** Top view of the fabricated first-generation Sierpiński fractal Kagome photonic crystal; the upper-left inset shows an enlarged view of 2×2 unit cells (the red dashed rhomboid represents a unit cell). **c** Top view of the fabricated second-generation Sierpiński fractal Kagome photonic. The upper-left inset shows an enlarged view of the unit cells. **d-f** Measured (colormaps) and simulated (white solid lines) bulk band structures of the samples in (**a-c**), respectively.

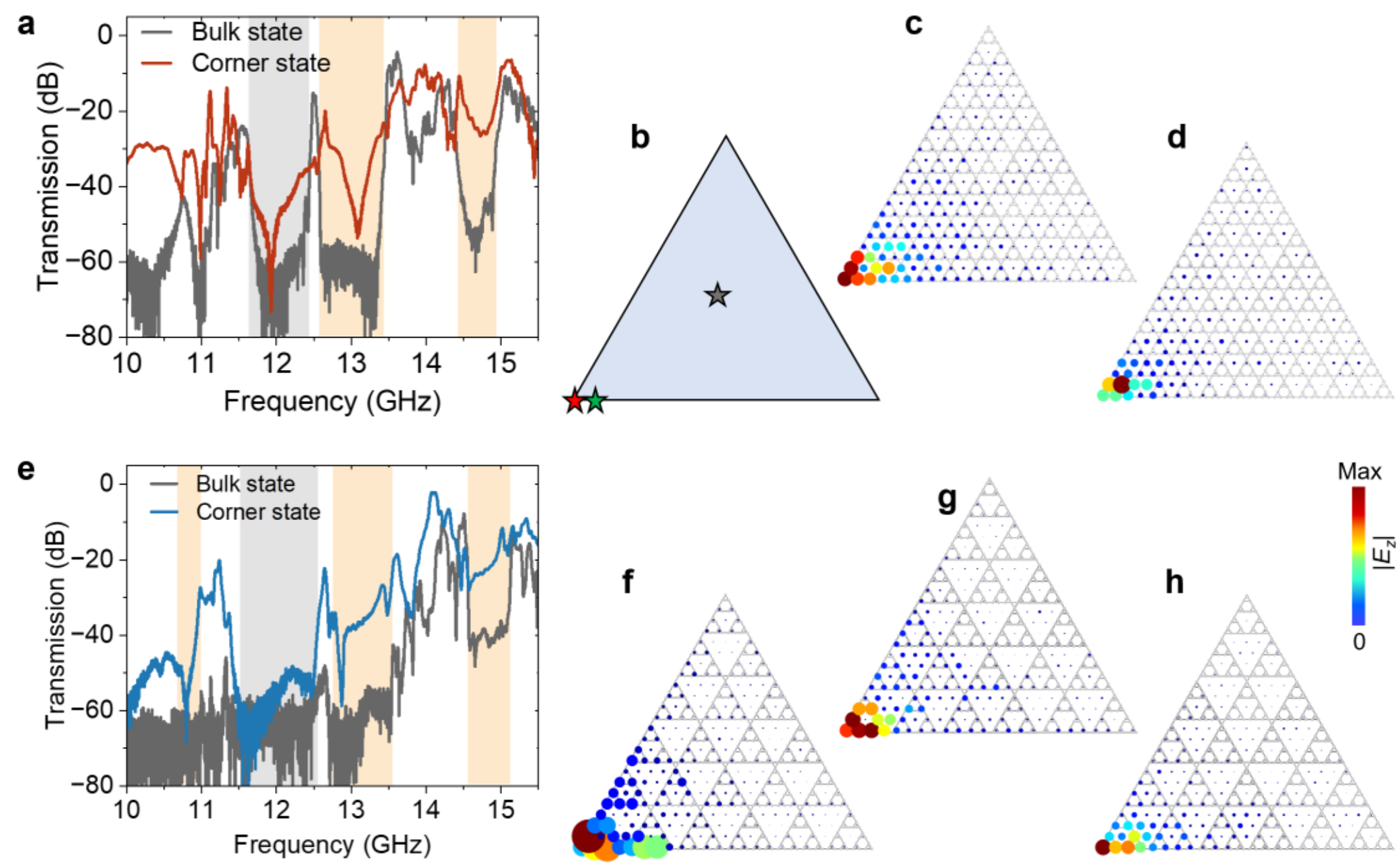

**Fig. 4 | Experimental observation of the fractality-induced photonic topological corner states**. **a** Measured transmission spectra of the corner (red line) and bulk (grey line) states for the first-generation Sierpiński fractal Kagome photonic crystal. The yellow (grey) regions represent topological (trivial) band gaps. **b** Experimental setup for the measurement. The green (red and grey) star represents the point source (probe). **c-d** Measured electric field distributions of the topological corner states at 13.4 GHz and 14.6 GHz, respectively. **e** Measured transmission spectra of the corner (blue line) and bulk (grey line) states for the second-generation Sierpiński fractal Kagome photonic crystal. **f-h** Measured electric field distributions of the topological corner states at 10.85 GHz, 13.4 GHz, and 14.8 GHz, respectively.